\documentclass[%
superscriptaddress,
 amsmath,amssymb,
 aps,
pre,
]{revtex4-2}

\usepackage{siunitx}
\usepackage{graphicx}
\usepackage{dcolumn}
\usepackage{bm}
\begin{document}
\preprint{APS/123-QED}
\title{Effects of particle elongation on dense granular flows down a rough inclined plane}

\author{Jixiong Liu}
\affiliation{%
Institute for Ocean Engineering, Shenzhen International Graduate School, Tsinghua University, Shenzhen, China}%
\author{Lu Jing}%
 \altaffiliation[Author to whom correspondence should be addressed: ]{lujing@sz.tsinghua.edu.cn}
\affiliation{%
Institute for Ocean Engineering, Shenzhen International Graduate School, Tsinghua University, Shenzhen, China}%
\affiliation{State Key Laboratory of Hydroscience and Engineering, Department of Hydraulic Engineering, Tsinghua University, Beijing, China}%
\affiliation{State Key Laboratory of Hydraulics and Mountain River Engineering, Sichuan University, Chengdu, China}%
\author{Thomas P\"ahtz}%
\affiliation{Institute of Port, Coastal and Offshore Engineering, Ocean College, Zhejiang University, Zhoushan, China
}%

\author{Yifei Cui}%
\affiliation{State Key Laboratory of Hydroscience and Engineering, Department of Hydraulic Engineering, Tsinghua University, Beijing, China}%
\affiliation{Key Laboratory of Hydrosphere Sciences of the Ministry of Water Resources, Beijing, China}%

\author{Gordon G. D. Zhou}%
\affiliation{Key Laboratory of Mountain Hazards and Earth Surface Process / Institute of Mountain Hazards and Environment, Chinese Academy of Sciences, Chengdu, China
}%

\author{Xudong Fu}%
\affiliation{State Key Laboratory of Hydroscience and Engineering, Department of Hydraulic Engineering, Tsinghua University, Beijing, China}%
\affiliation{Key Laboratory of Hydrosphere Sciences of the Ministry of Water Resources, Beijing, China}%

\date{\today}

\begin{abstract}
Granular materials in nature are nearly always non-spherical, but particle shape effects in granular flow remain largely elusive. This study uses discrete element method simulations to investigate how elongated particle shapes affect the mobility of dense granular flows down a rough incline. For a range of systematically varied particle length-to-diameter aspect ratios (AR), we run simulations with various flow thicknesses $h$ and slope angles $\theta$ to extract the well-known $h_\textrm{stop}(\theta)$ curves (below which the flow ceases) and the $Fr$-$h/h_\textrm{stop}$ relations following Pouliquen's approach, where $Fr=u/\sqrt{gh}$ is the Froude number, $u$ is the mean flow velocity, and $g$ is the gravitational acceleration. The slope $\beta$ of the $Fr$-$h/h_\textrm{stop}$ relations shows an intriguing S-shaped dependence on AR, with two plateaus at small and large AR, respectively, transitioning with a sharp increase. We understand this S-shaped dependence by examining statistics of particle orientation, alignment, and hindered rotation. We find that the rotation ability of weakly elongated particles ($\textrm{AR}\lesssim1.3$) remains similar to spheres, leading to the first plateau in the $\beta$-AR relation, whereas the effects of particle orientation saturates beyond $\textrm{AR}\approx2.0$, explaining the second plateau. An empirical sigmoidal function is proposed to capture this non-linear dependence. The findings are expected to enhance our understanding of how particle shape affects the flow of granular materials from both the flow- and particle-scale perspectives.
\end{abstract}

\maketitle
\section{\label{INTRODUCTION}INTRODUCTION}

Granular flow is ubiquitous in geophysical and industrial applications ranging from landslides, sediment transport, and volcanic ash dispersion to handling of bulk materials such as grains, pharmaceuticals, and coals~\cite{forterre2003long,tahmasebi2023state,weinhart2012closure,pouliquen2002friction}. Granular materials in real-world applications have complicated particle shapes~\cite{tahmasebi2023state}. However, despite extensive research into the fundamental mechanics of granular flow with spherical particles, our understanding of how particle shape affects the granular flow dynamics remains surprisingly inadequate. Recent studies indicate that particle shape can significantly affect various aspects of granular flow, including segregation~\cite{cunez2024particle,jones2021predicting}, fluid-driven transport~\cite{deal2023grain,qian2024numerical,cassel2021bedload}, rheology~\cite{borzsonyi2012orientational,hao2023rheology,mandal2018study}, and geomechanical behaviors~\cite{nie2024comprehensive,qian2018role}, but a general approach to considering the particle shape effects in state-of-the-art granular flow continuum models is still missing.

Granular flow down an inclined plane is a common model case for studying the dynamics of flowing granular materials~\cite{gdr2004dense,forterre2008flows}. \citet{pouliquen1999scaling} proposed a scaling law (also known as Pouliquen's flow rule) by investigating the average flow velocity ($u$) of the granular layer for various angles of inclination ($\theta$) and flow thicknesses ($h$). For spherical glass beads, it is found that the relation between the Froude number $Fr=u/\sqrt{gh}$ and the normalized flow thickness $h/h_\textrm{stop}$ can collapse all data onto a single master curve, where $g$ is the gravitational acceleration and $h_\textrm{stop}$ is a curve in the ($h$, $\theta$) parameter space delimiting no-flow and steady flow regimes. Follow-up experiments with sands (irregularly-shaped particles) showed that the $Fr$-$h/h_\textrm{stop}$ scaling holds, but the curve has a much steeper slope and does not pass through the origin, hinting at different mechanisms associated with material properties and particle shapes~\cite{forterre2003long}. In later research, \citet{borzsonyi2007flow} tested a range of materials, including glass beads, sands, and copper particles, and various particle sizes and bottom roughness conditions. Data collapse appears to be non-trivial, and they proposed to modify the $Fr$-$h/h_\textrm{stop}$ into a $Fr$-$h\tan ^2\theta/(h_\textrm{stop}\tan^2\theta_1)$ scaling for glass beads and sands, where $\theta_1$ is the dynamic angle of repose extracted from the $h_\textrm{stop}$ curve. However, large deviations from the latter relation occurred for copper particles with significantly non-rounded particle shapes. In particular, microscopic images of the copper particles showed a wide range of variations in the particle elongation and angularity, which can introduce additional micro-mechanical controls on the flow mobility compared with rounded particles. Particle elongation, roundness, angularity, and other shape factors of natural granular materials vary simultaneously, making it difficult to isolate different shape-related mechanisms from one another in physical experiments~\cite{zhao2023role}.

Here, we computationally investigate the effects of particle elongation in granular flows within the framework of Pouliquen's flow rule. We use the discrete element method (DEM) to simulate granular flows down a rough incline with systematically varied particle length-to-diameter aspect ratio, AR, and explore how AR affects the $Fr$-$h/h_\textrm{stop}$ relation at both the flow- and particle levels. Section \ref{METHODOLOGY} provides a detailed description of the DEM model setup and simulation protocols. In Section \ref{RESULTS}, we present the simulation results of typical flow behaviors for both spherical and elongated particles, followed by an analysis of the quantitative influence of AR on the characteristics of the $h_\textrm{stop}$ curve and $Fr$-$h/h_\textrm{stop}$ relations. Subsequently, we explore the microscopic mechanisms of the influence of AR in terms of particle alignment, orientation, and rotation, which provides insights into the complex interplay between the particle shape and granular flow dynamics. In Section \ref{Discussion}, we discuss the general implications of our findings and provide an empirical model for describing the $Fr$-$h/h_\textrm{stop}$ relation for elongated particles. Conclusions are drawn in Section \ref{Conclusions}.

\section{\label{METHODOLOGY}SIMULATION METHODOLOGY}
\subsection{\label{Numerical_setup}Numerical setup}
\begin{figure*}[htbp]
\includegraphics[width=0.9\textwidth]{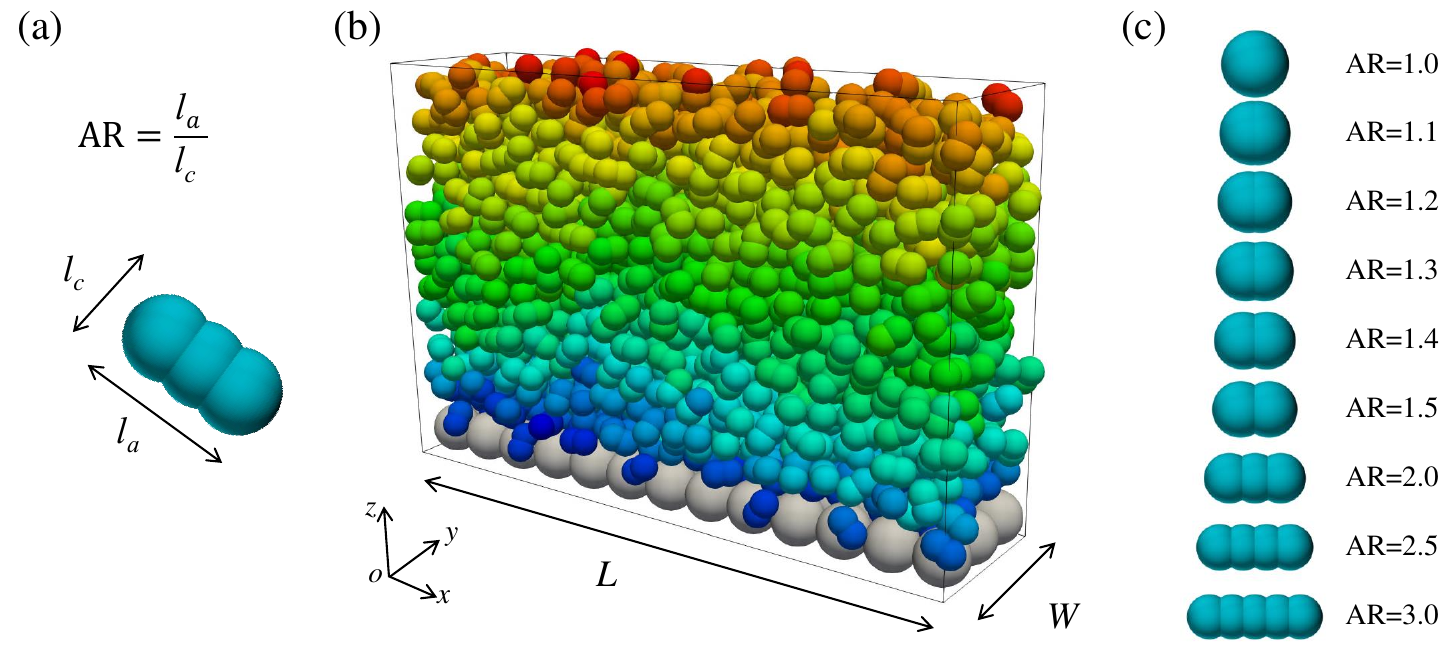} 
\caption{\label{fig1} Setup and primary notation. (a) Geometry of a typical elongated particle. (b) A snapshot of the DEM simulation. (c) Elongated particles with varying AR.}
\end{figure*}

The setup of our DEM simulations is illustrated in Fig.~\ref{fig1}. Most simulations are run for identical particles of volume-equivalent diameter $d$, modelled as clumps of spheres of a given preset aspect ratio AR, with their major and minor axes denoted by $l_a$ and $l_c$, respectively. Exceptions are runs for spherical particles ($\textrm{AR}=1$), for which a small polydispersity ($\pm0.1d$) is superimposed to avoid crystallization. Note that using clumped spheres to represent rod-like particles can introduce the so-called particle corrugation effect~\citep{guo2012numerical}, which is evaluated in Appendix~\ref{app:corrugation}. The computational domain is an open rectangular box of length $L$, width $W$, and no top. Periodic boundaries are imposed in the $x$- and $y$-directions to obtain a spatially homogeneous granular flows down an inclined plane without sidewalls. We choose $L=20l_a$ and $W=5l_a$ because these are about the smallest values above which the simulation results are insensitive to the domain size~\cite{pol2023unified} (see Appendix~\ref{app:domainsize}, Fig.~\ref{fig9}). We vary the initial thickness $H$ of the granular packing (by means of ``sand pluviation'') and the angle of inclination $\theta$ (by tilting the gravitational acceleration $g$) as two major flow parameters. A layer of regularly packed spheres of diameter $1.6d$ are fixed on the bottom to ensure a sufficiently large bottom roughness, required to avoid velocity slipping~\cite{silbert2001granular,weinhart2012closure,jing_characterization_2016}. The spacing between these bottom particles is adjusted such that they distribute evenly along $L$ and $W$ to avoid discontinuities at the periodic boundaries.

A linear spring-dashpot contact model is employed, which generates similar results as using Hertz model for dense granular flows as long as the particle stiffness is sufficiently high~\cite{silbert2001granular,weinhart2012closure}. The normal and tangential contact forces are computed in the same way as described by \citet{hao2023rheology}. We use a normal stiffness of $k_n=7.5\times 10^5 mg/d$, where $m$ is the particle mass. For particles of $d=1$~\si{mm} and $\rho=2650$~\si{kg/m^3}, and $g=9.81$~\si{m/s^2}, the corresponding particle stiffness is $k_n\approx10^4$~\si{N/m} and an equivalent Young's modulus can be estimated as $E=2k_n/d\approx2\times10^7$~\si{Pa}, according to \citet{hao2023rheology}. This choice is greater than the value of $2\times10^5 mg/d$ used by \citet{silbert2001granular}, with the latter being a common choice in the granular flow community~\cite{hidalgo2018rheological,thornton2012frictional,mandal2018study}. Our choice is made since further increasing $k_n$ has a minor influence on the results [Appendix~\ref{app:parameters} Fig.~\ref{fig10}(a)] but significantly reduces the computational efficiency. The tangential stiffness $k_t$ generally does not affect the results [see Appendix~\ref{app:parameters} Fig.~\ref{fig10}(b)] and is set to be equal to $k_n$. The normal restitution and tangential friction coefficients are $e=0.56$ and $\mu=0.5$, respectively, among common choices in DEM simulations of dense granular flows [see Appendix~\ref{app:parameters} Figs.~\ref{fig10}(c,d) for sensitivity analysis]. The timestep is set to be $\Delta t=6.26\times 10^{-5}\sqrt{d/g}$.

\subsection{\label{measurements}Simulation protocols and measurements}
For all aspect ratios $\textrm{AR}$, we conduct numerical simulations for two scenarios. The first scenario aims to determine the curve $h_\textrm{stop}(\theta)$, the critical thickness of the granular material below which the flow stops for a given inclination angle $\theta$ (Pouliquen, 1999). For a given value of $h_\textrm{stop}$ (i.e., a given initial layer thickness $H$), we run multiple test simulations to determine a critical value of $\theta$ below which a flow ceases. Practically, each simulation is initiated by accelerating the granular layer for a duration of about $16\sqrt{d/g}$ at $\theta=60^\circ$ and then the slope is lowered to a designated $\theta$ to deterimine whether the flow can be sustained. The critical stopping angle is defined as the mean between the smallest and largest slope angles at which a flow can and cannot be sustained, respectively. This procedure yields a pair of $h_\textrm{stop}$ and $\theta$ values and is repeated until a satisfactory $h_\textrm{stop}(\theta)$ curve is found. The second scenario examines the steady flow behavior of the granular material by systematically varying both $\theta$ and $H$ (Table~\ref{tab:table3}). Note that for different AR, the ranges of $\theta$ which allows a steady flow to be reached are different. To avoid instabilities of very thin layers~\cite{silbert2003granular}, $H$ is kept well above the $h_\textrm{stop}(\theta)$ curve. After a steady and uniform flow state has been reached, two time-averaged parameters are extracted: the mean particle velocity $u$ and the flow layer thickness $h$. Note that $h$ is typically greater than $H$ due to shear dilation and is calculated as twice the center of mass of the flowing particles, $h=2\sum_0^n{z_i}/n$, where $n$ is the number of particles and $z_i$ the vertical position of particle $i$. We also extract layer- and particle-wise information for necessary discussions in the subsequent analysis.

\begin{table}[h]
\caption{\label{tab:table3} Simulation conditions of all steady flow tests. }
\begin{ruledtabular}
\begin{tabular}{lcr}
\mbox{AR}&\mbox{$\theta$ (°)}&\mbox{$H/d$}\\
\hline
1.0 & 23, 24, 25, 26, 27 & 10, 15, 20, 25, 30, 35, 40\\
1.1 & 24, 25, 26, 27, 28 & 10, 15, 20, 25, 30, 35, 40\\
1.2 & 25, 26, 27, 28, 29 & 10, 15, 20, 25, 30, 35, 40\\
1.3 & 28, 29, 30, 31, 32 & 10, 15, 20, 25, 30, 35, 40\\
1.4 & 29, 30, 31, 32, 33 & 10, 15, 20, 25, 30, 35, 40\\
1.5 & 29, 30, 31, 32, 33 & 10, 15, 20, 25, 30, 35, 40\\
2.0 & 30, 31, 32, 33, 34 & 10, 15, 20, 25, 30, 35, 40\\
2.5 & 31, 32, 33, 34, 35 & 10, 15, 20, 25, 30, 35, 40\\
3.0 & 31, 32, 33, 34, 35 & 10, 15, 20, 25, 30, 35, 40\\
\end{tabular}
\end{ruledtabular}
\end{table}

\section{\label{RESULTS}SIMULATION RESULTS}
\subsection{Typical behaviors of spherical and elongated particles}

\begin{figure}[htbp]
\includegraphics[width=0.9\textwidth]{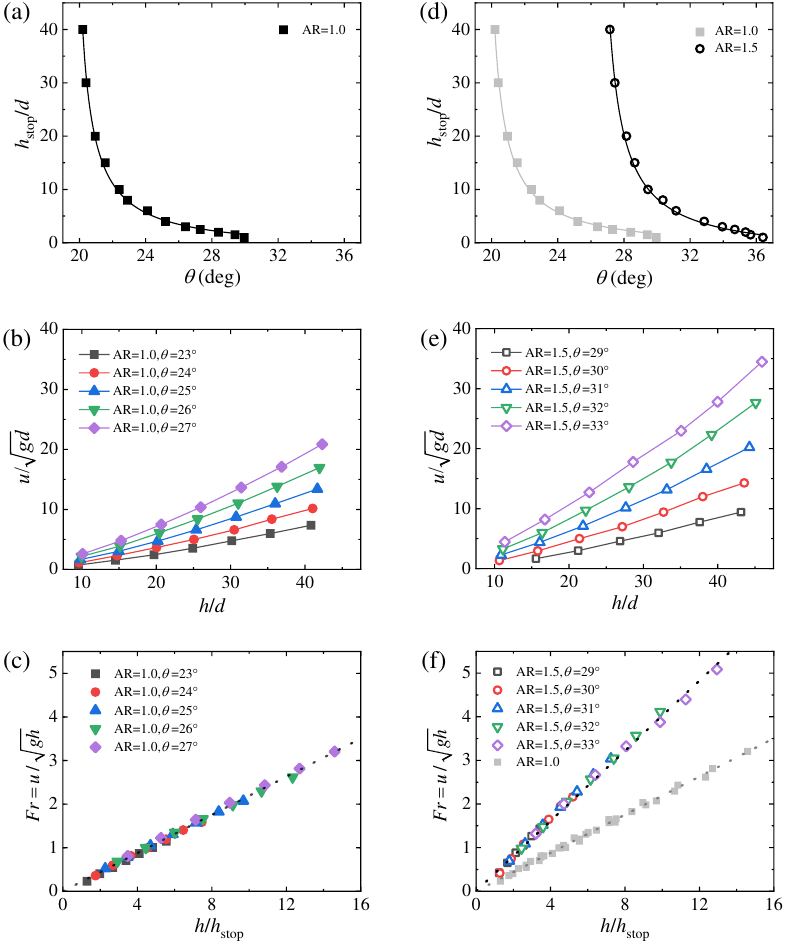}
\caption{\label{fig2}Typical flow behaviors of spherical ($\textrm{AR}=1$) and elongated ($\textrm{AR}=1.5$) particles. (a) $h_\textrm{stop}$ as a function of $\theta$, (b) $u/\sqrt{gd}$ as a function of $h/d$, and (c) Froude number $Fr=u/\sqrt{gh}$ as a function of $h/h_\textrm{stop}$ for $\textrm{AR}=1$. (d-f) same as (a-c) but with $\textrm{AR}=1.5$. Gray data points in (d) and (f) are $\textrm{AR}=1$ results for reference. Curves in (a) and (d) are fits to Eq.~(\ref{eq1}), while straight lines in (c) and (f) follow Eq.~(\ref{eq2}); fitting parameters can be found in Table~\ref{tab:fitpa}.}
\end{figure}

Figure~\ref{fig2} presents typical results for spherical ($\textrm{AR}=1$) and elongated particles ($\textrm{AR}=1.5$). It can be seen from Fig.~\ref{fig2}(a) that $h_\textrm{stop}/d$ decreases as $\theta$ increases, indicating that a thinner granular layer requires a greater inclination angle to flow. This relationship is typically fitted to the following expression~\cite{pouliquen1999scaling,forterre2003long,weinhart2012closure}: 
\begin{equation}\label{eq1}
\frac{h_\textrm{stop}}{d}=A\frac{\tan \theta _2-\tan \theta}{\tan \theta -\tan \theta _1}, 
\end{equation}
where $\theta_1$ is the angle towards which $h_\textrm{stop}(\theta)$ diverges and can be regarded as the ``dynamic angle of repose'' of the granular material, $\theta_2$ is the angle where $h_\textrm{stop}(\theta)$ vanishes, and $A$ is a dimensionless parameter controlling the shape of the $h_\textrm{stop}(\theta)$ curve.

Figure~\ref{fig2}(b) illustrates the relationship between the dimensionless mean velocity $u/\sqrt{gd}$ and the dimensionless flow thickness $h/d$ for different $\theta$, which shows that $u/\sqrt{gd}$ increases with $h/d$ and $\theta$. We follow Pouliquen's approach~\cite{pouliquen1999scaling} to rescale $h$ by $h_\textrm{stop}$ and $u$ by $\sqrt{gh}$. As found in previous work~\cite{pouliquen1999scaling,silbert2003granular,weinhart2012closure}, all data for spherical particles ($\textrm{AR}=1$) collapse onto a single master curve [Fig.~\ref{fig2}(c)]. The master curve can be fitted to a linear function passing through the origin:
\begin{equation}\label{eq2}
\frac{u}{\sqrt{gh}}=\beta \frac{h}{h_\textrm{stop}},
\end{equation}
where $\beta$ is a dimensionless constant (slope of the fitted line) and is a key parameter of Pouliquen's flow rule. 

Figures \ref{fig2}(d--f) present results for elongated particles with $\textrm{AR}=1.5$. The $h_\textrm{stop}(\theta)$ curve shifts to the right significantly [Fig.~\ref{fig2}(d)], indicating that elongated particles are much more difficult to flow. Comparing Fig.~\ref{fig2}(b) and Fig.~\ref{fig2}(e), the dependence of $u/\sqrt{gd}$ on $h/d$ and $\theta$ for elongated particles is similar to that of spherical particles, although the range of $u/\sqrt{gd}$ is generally larger for $\textrm{AR}=1.5$ due to the higher values of $\theta$ needed in these simulations. In Fig.~\ref{fig2}(f), data collapse is achieved for $\textrm{AR}=1.5$ but onto a significantly different curve (which also tends to pass through the origin) than the master curve of $\textrm{AR}=1$. Fitting the data to Eq.~(\ref{eq2}) yields a value of $\beta$ for elongated particles much higher than that of spheres, although in physical units elongated particles exhibit less mobility given a similar set of $h$ and $\theta$ [comparing Figs.~\ref{fig2}(b) and (e)]. Figure~\ref{fig2}(f) also shows that although Pouliquen's scaling approach can be applied to collapse the data of elongated particles for a wide range of $h/d$ and $\theta$ (which has not been systematically confirmed before), collapsing data for both spherical and elongated particles onto a single curve seems to be non-trivial, which hints at different flow mechanisms for both types of particles. Our next focuses are to examine how $\beta$ varies with AR and to explore the underlying micromechanical origin of such shape effects.

\begin{figure*}[htbp]
\includegraphics[width=0.9\textwidth]{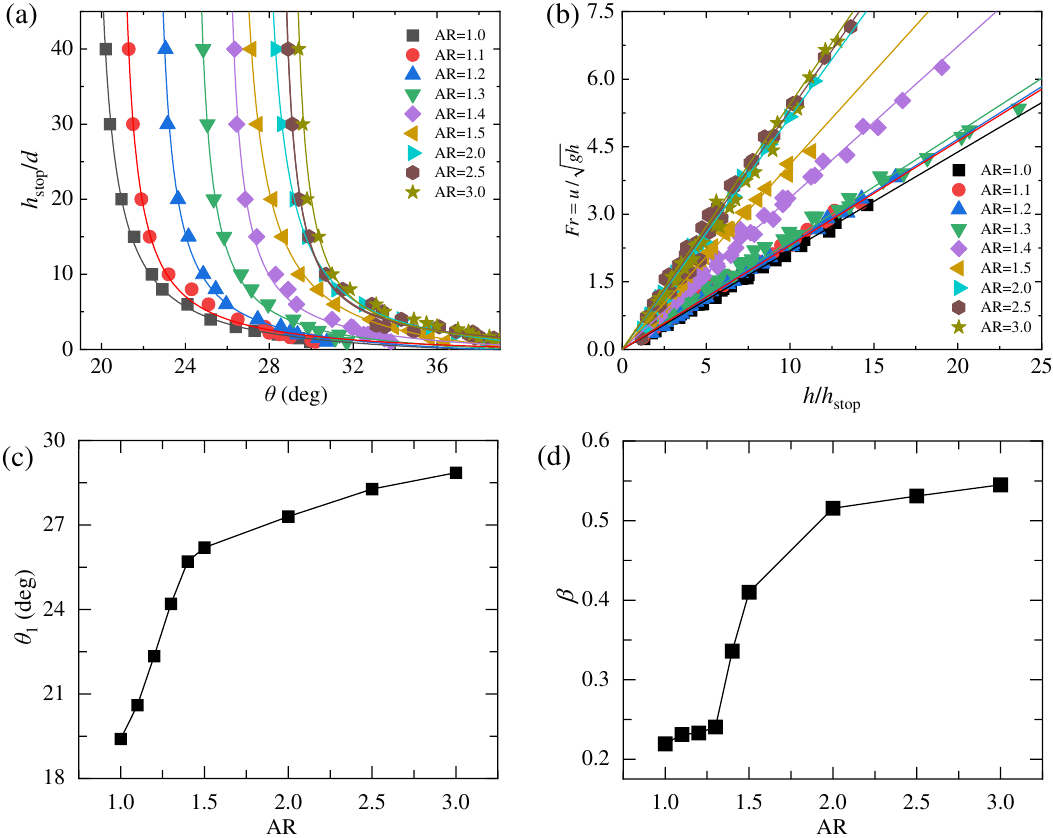}
\caption{\label{fig3} Flow behavior with varying particle aspect ratio (AR). (a) $h_\textrm{stop}$ as a function of $\theta $. (b) Froude number $u/\sqrt{gh}$ as a function of $h/h_\textrm{stop}$. (c) Fitting parameter $\theta_1 $ and (d) fitting parameter $\beta$ as a function of AR. Curves in (a) and (b) are fits to Eq.~(\ref{eq1}) and Eq.~(\ref{eq2}), respectively, with fitting parameters listed in Table~\ref{tab:fitpa}.}
\end{figure*}

\subsection{Varying particle aspect ratios AR}
For varying aspect ratios $1\leqslant\textrm{AR}\leqslant3$, Fig.~\ref{fig3}(a) shows that, as AR increases, the $h_\textrm{stop}$ curve generally shifts to the right but the trend tends to saturate as $\textrm{AR}$ reaches about $2$.  In other words, at a given $\theta$, particles with a greater AR require a thicker flow layer $h$ to maintain a steady fully developed flow. The solid curves in Fig.~\ref{fig3}(a) represent the fits to the $h_\textrm{stop}$ data using Eq.~(\ref{eq1}) and the corresponding fitting parameters are listed in Table~\ref{tab:fitpa}. Note that the fitting errors for parameters $\theta_2$ and $A$ are typically larger than for $\theta_1$ due to natural uncertainties at $h_\textrm{stop}/d\sim O(1)$, which nevertheless does not affect our data analysis below. Interestingly, $\theta_1$ depends systematically on AR, but the variations of $\theta_2$ with AR fall roughly within the range of fitting errors. To better correlate the change of the $h_\textrm{stop}$ curves with AR, we plot the dynamic angle of repose $\theta_1$ against AR in Fig.~\ref{fig3}(c). A non-linear increasing trend is observed, which tends to saturate beyond $\textrm{AR}=1.5$. This non-linear dependence of the angle of repose with the particle aspect ratio is similar to previous results~\cite{borzsonyi2007flow} and implies a change of how particle elongation affects the flow mobility, as discussed below in Section IV. 

\begin{table}[h]
\caption{\label{tab:fitpa} Fitting parameters for Fig.~\ref{fig3}. }
\begin{ruledtabular}
\begin{tabular}{lcccr}
\mbox{AR} & \mbox{$\theta_1$ (°) (fitting error)} & \mbox{$\theta_2$ (°) (fitting error)} & \mbox{$A$ (fitting error)} & \mbox{$\beta$ (fitting error)} \\
\hline
1.0 & $19.40\enspace( \pm0.04)$ & $37.68\enspace( \pm3.06)$ & $1.65\enspace( \pm0.41)$ & $0.219\enspace( \pm0.002)$ \\ 
1.1 & $20.60\enspace( \pm0.04)$ & $43.74\enspace( \pm6.41)$ & $0.95\enspace( \pm0.40)$ & $0.231\enspace( \pm0.002)$ \\ 
1.2 & $22.34\enspace( \pm0.07)$ & $38.39\enspace( \pm5.93)$ & $1.51\enspace( \pm0.84)$ & $0.233\enspace( \pm0.005)$ \\ 
1.3 & $24.20\enspace( \pm0.03)$ & $39.92\enspace( \pm3.46)$ & $1.43\enspace( \pm0.45)$ & $0.240\enspace( \pm0.001)$ \\ 
1.4 & $25.70\enspace( \pm0.07)$ & $44.71\enspace( \pm8.18)$ & $1.09\enspace( \pm0.53)$ & $0.336\enspace( \pm0.001)$ \\ 
1.5 & $26.20\enspace( \pm0.04)$ & $40.80\enspace( \pm0.96)$ & $2.71\enspace( \pm0.29)$ & $0.411\enspace( \pm0.001)$ \\ 
2.0 & $27.30\enspace( \pm0.03)$ & $44.77\enspace( \pm1.60)$ & $1.96\enspace( \pm0.29)$ & $0.516\enspace( \pm0.001)$ \\ 
2.5 & $28.28\enspace( \pm0.02)$ & $60.99\enspace( \pm8.16)$ & $0.44\enspace( \pm0.23)$ & $0.532\enspace( \pm0.003)$ \\ 
3.0 & $28.86\enspace( \pm0.04)$ & $53.48\enspace( \pm7.61)$ & $0.68\enspace( \pm0.36)$ &$0.541\enspace(\pm0.002)$ \\ 
\end{tabular}
\end{ruledtabular}
\end{table}
The $u/\sqrt{gh}$-$h/h_\textrm{stop}$ relationships for varying AR are presented in Fig.~\ref{fig3}(b). Similar to $\textrm{AR}=1$ or $1.5$, the results (with varying $H$ and $\theta$ as listed in Table~\ref{tab:table3}) for each AR fall on a single master curve passing through the origin, indicating that Pouliquen's scaling holds for each AR. However, this does not collapse the data altogether but leads to two main clusters of data for $1\leqslant\textrm{AR}\lesssim1.3$ and $2\lesssim\textrm{AR}\leqslant3$, respectively, with the slope of the data generally increases with AR. To show this trend more clearly, we fit each dataset to Eq.~\ref{eq2} (see Table~\ref{tab:fitpa} for fitted values of the slope $\beta$) and plot $\beta$ against AR in Fig.~\ref{fig3}(d). An S-shaped curve is observed: for $\textrm{AR}\lesssim1.3$, $\beta$ increases slowly with the increase of AR, indicating that the flow mobility is insignificantly affected by particle elongation in this regime; for $1.3\lesssim\textrm{AR}\leqslant2$, $\beta$ exhibits a sharp increase from around $0.24$ to $0.516$, which signifies a fundamental change of the flow dynamics of elongated particles compared to sphere-like particles; finally, as AR increases beyond $2$, the increase of $\beta$ slows down, indicating that the shape effects on the flow mobility tend to saturate in this regime. To understand the physics of this three-regime behavior, we explore microscopic aspects of the flowing elongated particles below.

\subsection{Microscopic analysis}
\subsubsection{\label{sec:alignment}Particle alignment and orientation}

\begin{figure*}[htbp]
\centering
\includegraphics[width=0.9\textwidth]{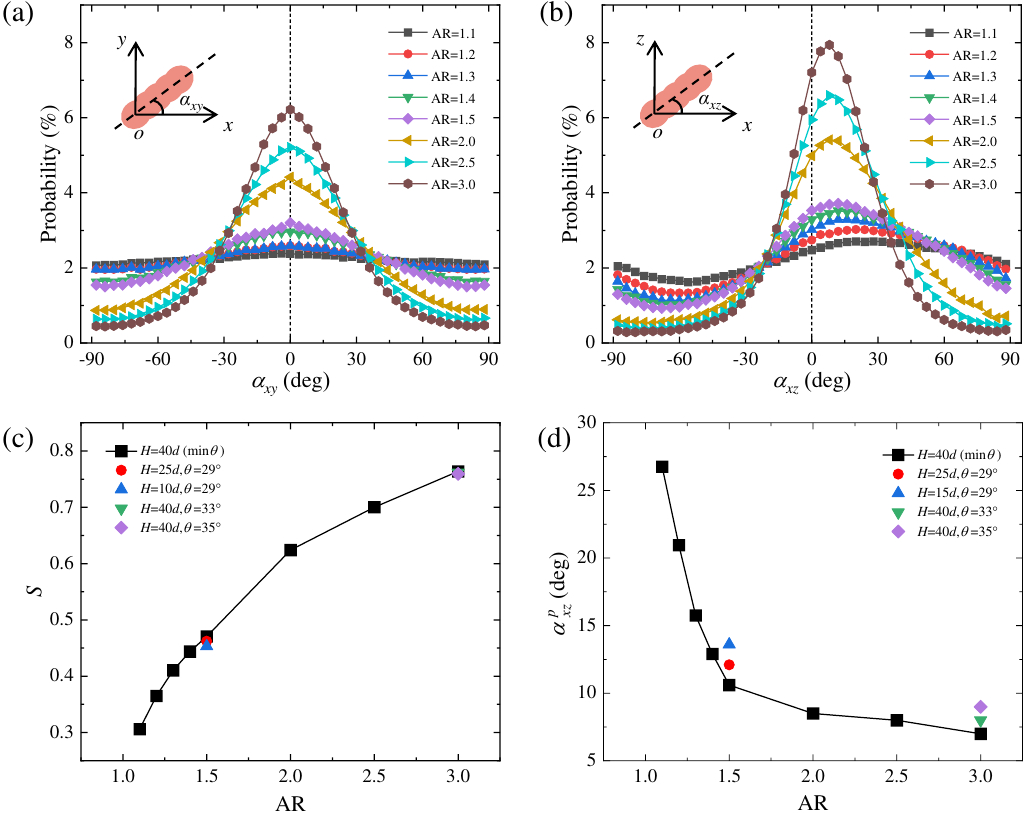}
\caption{\label{fig4} Particle alignment and orientation for AR ranging from 1.1 to 3 (with $H=40d$ and the corresponding minimum $\theta$ listed in Table~\ref{tab:table3}). (a) Probability distribution of particle orientation in the $xy$ plane $\alpha_{xy}$ and (b) in the $xz$ plane $\alpha_{xz}$. (c) The order parameter $S$ and (d) the preferred orientation angle $\alpha_{xz}^{p}$ as a function of AR.}
\end{figure*}

Particle alignment along a preferential orientation is perhaps the most straightforward effect of elongated particle shapes in granular flows~\cite{guo2015discrete,guo2013granular,nadler2018kinematic}. To characterize the statistics of the particle orientation, particles are projected onto the $xy$ (top view) and $xz$ (side view) planes to define angles $\alpha_{xy}$ and $\alpha_{xz}$, respectively (see insets of Fig.~\ref{fig4}). As we confirm that the particle orientation statistics are insensitive to the flow depth at which measurements are taken, the statistics we report here are obtained over all particles  except those within $2d$ of the wall and free surface for each simulation~\cite{mandal2018study,pol2022kinematics,cai2021diffusion,borzsonyi2012orientational}. Moreover, since the orientation statistics are also insensitive to flow conditions $h$ and $\theta$ (see data below), we generally use one example simulation for each AR (with $h=40d$ and the corresponding minimum $\theta$ listed in Table~\ref{tab:table3}) to demonstrate the results. Figure~\ref{fig4}(a) shows that the probability distributions of $\alpha_{xy}$ are basically symmetric about $\alpha_{xy}=0$ for all AR and are narrower as AR increases. This means that more elongated particles tend to align better along the flow direction ($x$) in the $xy$ plane but without a sideways preference. By contrast, Fig.~\ref{fig4}(b) shows that the probability distributions of $\alpha_{xz}$ are skewed with mean values lying approximately between $0$ and $30^\circ$ (each curve remains nearly a normal distribution but is shifted to the right, hence the dips observed at around $-60^\circ$ for large AR~\cite{mandal2018study,pol2022kinematics}). A relevant study found similarly that in granular shear flows with large particle aspect ratios and high solid volume fractions, the angle $\alpha_{xz}$ predominantly falls within the range of 0 to $10^\circ$~\cite{guo2012numerical}. This indicates that, as AR increases, elongated particles tend to align themselves with a certain preferential orientation with respect to the flow direction in the $xz$ plane (pointing forward and slightly upward when viewed from the side). 

The analysis above shows that the statistics of $\alpha_{xz}$, including the degree of alignment (narrowness of the distribution) and the preferential orientation (mean value), are important microscopic descriptors of flowing elongated particles. To be more quantitative, in Fig.~\ref{fig4}(c), we introduce a parameter that characterizes the particle alignment in the system, namely, the orientation order parameter $S$~\cite{mandal2018study},
\begin{equation}\label{eq3}
S=\frac{3\left< \cos ^2\left( \alpha_{xz}-\alpha_{xz}^{avg} \right) \right> -1}{2},
\end{equation}
where $\alpha _{xz}$ is the angle between the longest axis of particles and the flow direction, $\alpha _{xz}^{avg}$ is the average orientation angle of all particles in a simulation, and $\left< \right>$ represents the ensemble average.

Figure~\ref{fig4}(c) shows that as AR increases, $S$ increases but with a decreasing rate, consistent with previous research~\cite{cai2021diffusion,mandal2018study}. Note that for $\textrm{AR}=1$, $S$ is undefined as a spherical particle does not have a definite long axis. For AR as small as $1.1$, a considerable degree of alignment ($S\approx0.3$) appears, meaning that even slightly elongated particles can develop a tendency to arrange themselves. This tendency of alignment increases gradually with AR, but a sharp transition similar to Fig.~\ref{fig3}(c) is not observed. These observations seem to indicate that particle alignment cannot explain the S-shaped dependence of $\beta$ on AR observed in Fig.~\ref{fig3}(c). Nevertheless, it is clear that, as the particle is sufficiently elongated ($\textrm{AR}\gtrsim2$), the tendency of particle alignment tends to saturate, partly explaining the second plateau of Fig.~\ref{fig3}(c). Finally, Fig.~\ref{fig4}(c) also include data for $\textrm{AR}=1.5$ and $3.0$ with various $H$ and $\theta$, showing that particle alignment is not much affected by flow conditions but mainly a function of AR in our simulations.

In Fig.~\ref{fig4}(d), we quantify the preferred orientation $\alpha_{xz}^{p}$ for different AR. This value describes the orientation at which particles are most likely to appear. It shows that $\alpha _{xz}^{p}$ first decreases sharply as AR increases and then this trend slows down significantly at around $\textrm{AR}=1.5$, similar to previous research~\cite{cai2021diffusion,mandal2018study,nagy2020flow}. The transition at $\textrm{AR}=1.5$ is similar to the trend of the $\beta$-AR plot in Fig.~\ref{fig3}, explaining the saturation of the shape effects on the flow mobility. Figure~\ref{fig4}(d) also includes data for $\textrm{AR}=1.5$ and $3.0$ with various $H$ and $\theta$, showing that the alignment is only slightly affected by the flow conditions.

The order parameter $S$ and the preferred orientation angle $\alpha_{xz}^{p}$ provide valuable insights into how elongated particle shapes affect flow mobility from a microscopic point of view. The specific alignment and orientation of particles can significantly influence the contact dynamics between particles, potentially leading to increased friction and hindered particle rotation\cite{guo2012numerical,guo2015discrete}. Consequently, it may result in a decrease in the flow mobility, consistent with \citet{guo2012numerical} that non-spherical particles convert some of their translational energy into rotational energy after collisions, resulting in increased energy dissipation. The trends observed in Figs.~\ref{fig4}(c) and (d) offer a plausible partial explanation for the observed trend of $\beta$ varying with AR [Fig.~\ref{fig3}(d)]. The correlation between the trends observed in $S$, $\alpha _{xz}^{p}$, and $\beta$ highlights the intricate relationship between particle shape, alignment, and flow mobility. The rapid initial changes in $\beta$ for $\textrm{AR}\gtrsim1.3$ can be directly linked to the corresponding changes in $S$ and $\alpha _{xz}^{p}$. However, when $\textrm{AR}\lesssim1.3$, the alignment and orientation for particles cannot explain the observation in Fig.~\ref{fig3}(d) that $\beta$ remains almost unchanged.

\subsubsection{\label{sec:angular velocity}Particle angular velocity}

\begin{figure*}[htbp]
\includegraphics[width=0.9\textwidth]{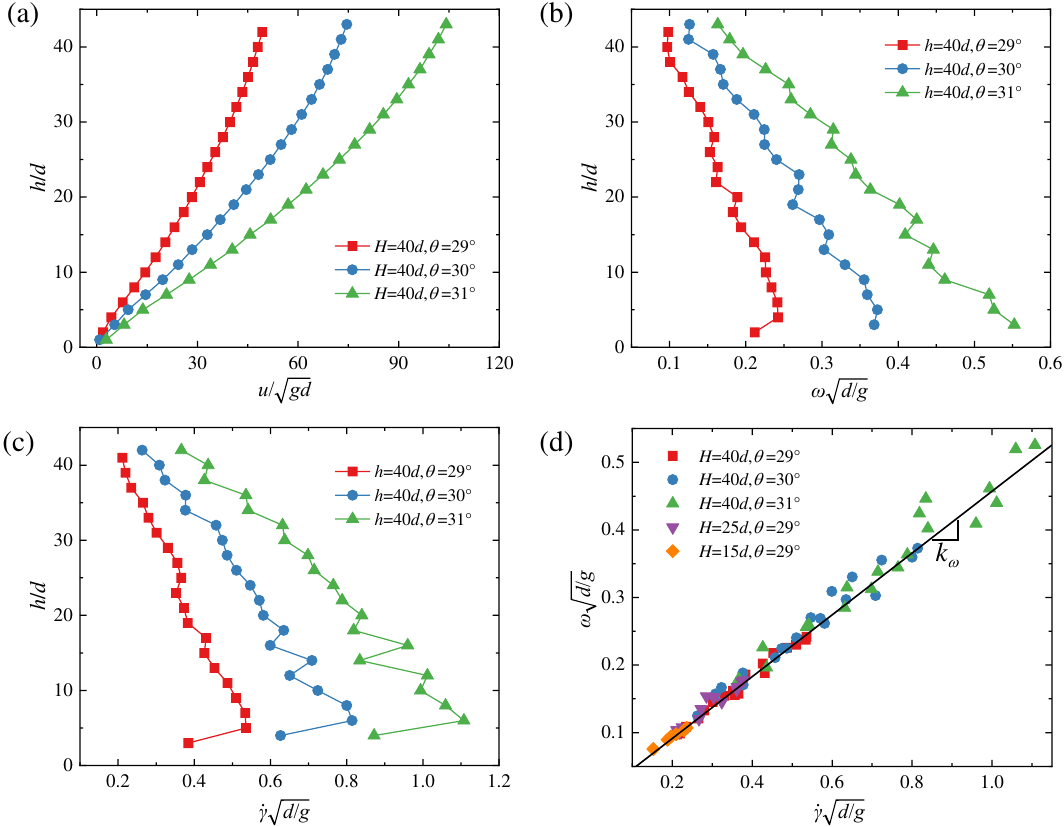}
\caption{\label{fig5} Layerwise measurements of flow kinematics for $\textrm{AR}=1.5$ and varying $H$ and $\theta$. (a) Particle velocity profile, (b) particle angular velocity profile, and (c) velocity gradient (shear rate) profile. (d) Non-dimensional angular velocity $\omega\sqrt{d/g}$ vs non-dimensional shear rate $\dot\gamma\sqrt{d/g}$ under different flow conditions.}
\end{figure*}

Particle angular velocity is a crucial parameter that reflects the ability of particles to rotate during the flow. To understand how AR affects the rotation dynamics of elongated particles (hence the flow mobility), we measure the mean rotational velocity of particles around the $y$-axis [see inset of Fig.~\ref{fig4}(b)], denoted by $\omega$, at a certain flow layer (with thickness $2d$). We then examine how $\omega$ depends on the flow kinematics (velocity and shear rate profiles) for various flow conditions. Note that the shear rate $\dot\gamma=\partial u/\partial z$ is determined from the velocity profile by dividing the velocity difference between layers by the layer thickness $2d$.  Using $\textrm{AR}=1.5$ as an example, Figs.~\ref{fig5}(a-c) present the flow-depth profiles of the dimensionless flow velocity $u/\sqrt{gd}$, rotational velocity $\omega\sqrt{d/g}$, and shear rate $\dot\gamma\sqrt{d/g}$ for three different flow conditions ($H=40d$ and $\theta=29^\circ$, $30^\circ$ and $31^\circ$). Understandably, $\omega\sqrt{d/g}$ is proportional to $\dot\gamma\sqrt{d/g}$ because the velocity difference between two adjacent layers drives particle rotation. Plotting $\omega\sqrt{d/g}$ against $\dot\gamma\sqrt{d/g}$ in Fig.~\ref{fig5}(d) leads to a nice linear correlation for five different flow conditions (with varying $H$ and $\theta$), noting that higher flow layer thicknesses and larger slope angles result in a wider range of shear rates and angular velocities. This result indicates that, for a given particle shape ($\textrm{AR}=1.5$ here), the mean rotational velocity at the flow level is, indeed, determined by the local shear rate of the flow. The slope of the plot in Fig.~\ref{fig5}(d), which we refer to as $k_\omega$, can be considered as a measure of the rotation ability of a particular granular material under shear conditions. 

\begin{figure*}[htbp]
\includegraphics[width=0.9\textwidth]{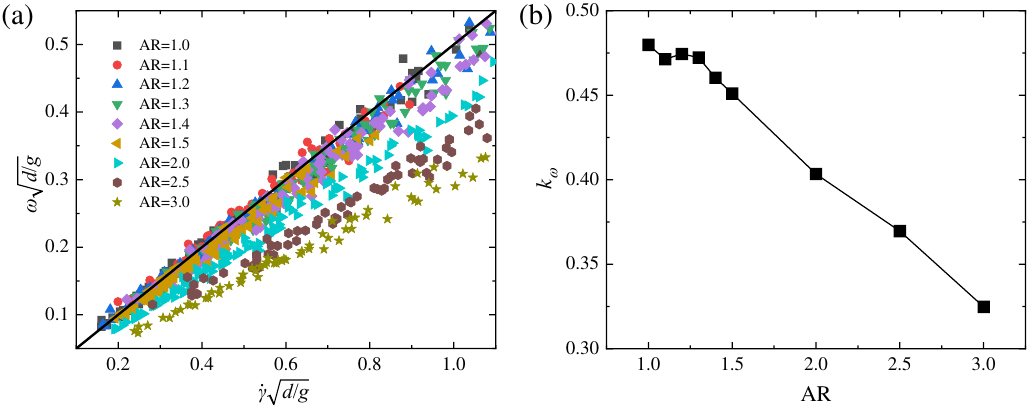}
\caption{\label{fig6}Rotational statistics for varying AR. (a) Non-dimensional angular velocity $\omega\sqrt{d/g}$ vs.\ non-dimensional shear rate $\dot\gamma\sqrt{d/g}$. The black line represents a fit to the data for $\textrm{AR} = 1$. (b) Particle rotation ability $k_\omega$ as a function of AR.}
\end{figure*}

To further investigate how AR affects the rotation ability of the particles, $k_\omega$, we show $\omega$-$\dot\gamma$ relations and the corresponding $k_\omega$ values for various AR in Fig.~\ref{fig6}. Note that, in Fig.~\ref{fig6}(a), all simulation conditions associated with a given AR are used together (indicated by the same color) to enhance statistics, which is justified as the $\omega$-$\dot\gamma$ relation is insensitive to flow conditions (Fig.~\ref{fig5}). The results in Fig.~\ref{fig6} reveal an interesting plateau regime for $\textrm{AR}\lesssim1.3$, including data for spherical particles ($\textrm{AR}=1$), which indicates that although the flow mobility (translational and rotational velocities) of slightly elongated particles differ significantly from spherical particles (see Figs.~\ref{fig2} and \ref{fig5}), the intrinsic rotation ability of the particles ($k_\omega$) remains essentially unchanged in this regime. Moreover, it is intriguing that the plateau value of $k_\omega$ for $\textrm{AR}\lesssim1.3$ is only slightly below $0.5$, as noted previously~\cite{mandal2018study}, and may be related to the ratio $0.5$ between vorticity and shear rate in fluid mechanics~\cite{batchelor1967introduction}; of course, this connection between granular flows and ordinary fluids warrants further investigation. Beyond the plateau regime of $\textrm{AR}\lesssim1.3$, Fig.~\ref{fig6} shows that $k_\omega$ decreases rapidly with AR, reflecting the fact that elongated particles experience greater resistance to rotation compared to their more spherical counterparts. This observation suggests the existence of a critical AR beyond which the elongated particle shape substantially restricts the rotation ability of the particles, and the threshold behavior in $k_\omega$ seems to explain the initial plateau in the $\beta$-AR plot in Fig.~\ref{fig3}(d).

\section{\label{Discussion}Discussion}
\subsection{Comparing with previous results}

\begin{figure*}[htbp]
\includegraphics[width=0.9\textwidth]{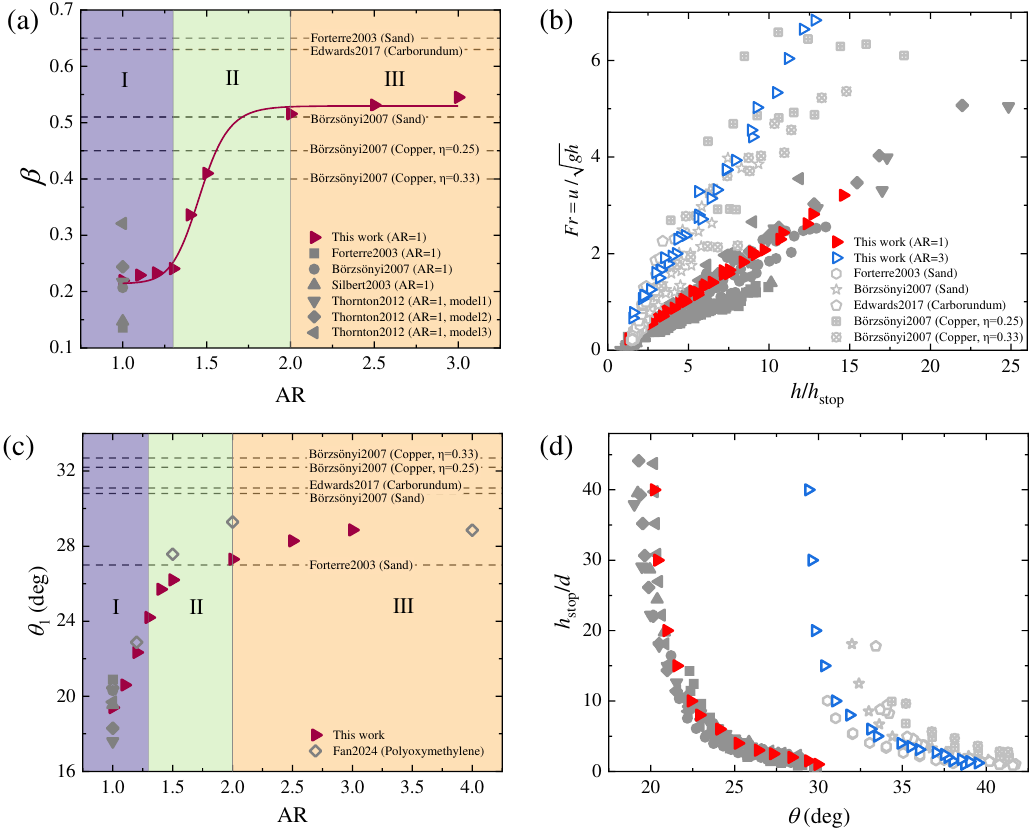}
\caption{\label{fig7} Comparison of our key results with data from the literature~\cite{borzsonyi2007flow,edwards2017formation,silbert2003granular,forterre2003long,thornton2012frictional,fan2024effect}. (a) Fitted parameter $\beta$ and (c) dynamic angle of repose $\theta_1$ as a function of AR. Gray symbols and dashed lines represent previous data from the literature, with detailed information presented in the text. (b) $Fr$-$h/h_{\text{stop}}$ relations and (d) $h_{\text{stop}}$ curves for data from the literature (gray symbols) comparing to our results (red and blue symbols).}
\end{figure*}

A key result above is the sigmoidal dependence of the flow rule parameter $\beta$ [Fig.~\ref{fig3}(d)], as well as the bi-linear dependence of the dynamic angle of repose $\theta_1$ [Fig.~\ref{fig3}(c)],  on the particle aspect ratio AR. To put our results into a more general context and discuss their broader implications, in Fig.~\ref{fig7} we compile and compare experimental and simulation data available from the literature, which deal with various materials including spheres (glass beads), elongated particles, sands, and Carborundum. Generally, results for $\text{AR}=1$ (spheres in DEM and glass beads in lab) and $\text{AR}=3$ (or realistic non-spherical particles in lab) form two data clusters in all plots of Fig.~\ref{fig7}, and our results of varying AR show the transition between the two clusters. Note that since realistic materials (sand, copper, and carborundum) do not have a definite AR (and they vary in other shape parameters), and the range of AR can vary widely encompassing $\textrm{AR} = 1 \sim 3$ in the literature~\cite{barrett1980shape,zhao20203d,fan2023improved,blott2008particle}, we show their results as horizontal dashed lines in Figs.~\ref{fig7}(a,c) to enable comparison with our results of varying AR.

More specifically, the dark gray solid points in Fig.~\ref{fig7} represent spherical particle data from \citet{forterre2003long} (0.5 mm glass beads) and \citet{borzsonyi2007flow} (0.51 mm glass beads) for experiments, and \citet{silbert2003granular} and \citet{thornton2012frictional} for simulations. \citet{silbert2003granular} used the Hertzian contact model with $\mu = 0.5$, $k_n = 2 \times 10^5 {mg}/{d}$, $k_t = 2/7k_n$, normal viscoelastic constants $\gamma_n = 50\sqrt{{g}/{d}}$, and tangential viscoelastic constants $\gamma_t = {\gamma_n}/{2}$. \citet{thornton2012frictional} used the same parameters but with $\gamma_t = \gamma_n$, and employed three contact models: spring-dashpot (model1), plastic (model2), and Hertzian (model3). For non-spherical particles (light gray hollow symbols), \citet{forterre2003long} used sand particles ($0.8 \pm 0.1$ mm), \citet{borzsonyi2007flow} used sand particles (0.4 mm) and copper particles ($0.16 \pm 0.03$ mm) with varying anisotropic dendritic shapes (indicated by apparent volume fractions $\eta = 0.25$ and $\eta  = 0.33$, respectively), and \citet{edwards2017formation} used carborundum particles $(0.28\sim0.35$ mm). Notably, Fig.~\ref{fig7}(c) presents experimental data from \citet{fan2024effect} using 7 mm Polyoxymethylene ellipsoids of varying AR, which closely match our numerical results of $\theta_1$. The slightly higher experimental results are likely due to wall effects in the quasi-2D cell~\cite{fan2024effect}. Despite the variations due to detailed differences in the numerical model and natural materials, it is clear that the two extreme ends of our data correspond well with realistic spheres (glass beads) and irregular materials. This supports the validity of our results and builds a connection (S-shaped transition) between the extremes cases with fine-tuned AR values. Future work should generate results for more different particle shapes (e.g., varying flatness and angularity) to further enrich this picture. Finally, we discuss our data in an alternative framework called improved Pouliquen-Jenkins flow rule~\cite{borzsonyi2007flow} in Appendix~\ref{app:PJFR}, which also shows excellent agreement between our results and previous data for various granular materials.

\subsection{Implications for modeling shallow granular flows}

Since the $\beta$ value is a key model parameter for shallow granular flow modeling in geophysical applications~\cite{pouliquen1999scaling,forterre2003long}, we establish an empirical model for $\beta$ in terms of AR. The sigmoidal trend in Fig. \ref{fig7}(a) can be captured by a logistic function fit, as demonstrated by the solid curve $\beta_\textrm{AR}$,

\begin{equation}\label{eq4}
\beta_\textrm{AR} = A_2 + \frac{A_1 - A_2}{1 + (\textrm{AR}/A_0)^p}
\end{equation}
where the fitted parameters are $A_0 = 1.46$, $A_1 = 0.22$, $A_2 = 0.53$, and $p = 17.83$. The $\beta$-AR curve can be divided into three regions: In the first region ($1.0 < \textrm{AR} \lesssim 1.3$), $\beta$ remains nearly unchanged. This indicates that although particle alignment already occurs (Fig.~\ref{fig4}), the unchanged ability of particle rotation in this small elongation region (Fig.~\ref{fig6}) seems to control the flow mobility. In the second region ($1.3 \lesssim \textrm{AR} \lesssim 2.0$), a sharp increase of $\beta$ is observed, corresponding to the decline of $k_\omega$ in Fig.~\ref{fig6}(b). Both mechanisms of rotation and alignment seem to contribute to this sharp transition. In the third region ($\textrm{AR} \gtrsim 2.0$), particle alignment effects saturate, leading to the second plateau, which suggests that the rotation ability in this region is small enough to be dominated by mechanisms related to particle alignment.

To show the potential use of Eq.~(\ref{eq4}), we present a set of governing equations of shallow granular flows that finds applications in geophysical mass flow modeling~\cite{savage1989motion,forterre2003long,forterre2008flows}:
\begin{equation}\label{shallow1}
\frac{\partial h}{\partial t}+\frac{\partial hu}{\partial x}=0,
\end{equation}
\begin{equation}\label{shallow2}
\\\rho\left(\frac{\partial hu}{\partial t}+\alpha\frac{\partial hu^2}{\partial x}\right)=\left(\tan\theta-\mu_b-K\frac{\partial h}{\partial x}\right)\rho gh\cos\theta,
\end{equation}
where $t$ is time, $h$ is the local flow thickness, $u$ is the depth-averaged flow velocity, $\rho$ is flow density, $\mu_b$ is basal friction, and $\alpha$ and $K$ are model parameters. Eqs.~(\ref{shallow1}) and (\ref{shallow2}) represent mass and momentum conservation equations, respectively. The basal friction term $\mu_b$ encapsulates the primary complex three-dimensional rheology of granular materials~\cite{forterre2003long,forterre2008flows}, as well as the particle shape effects on the flow mobility as discussed above. By combining Eqs.~(\ref{eq1}) and (\ref{eq2}), and considering the system to be at steady state ($\mu_b = \tan\theta$), we arrive at the following extended Pouliquen flow rule accounting for particle shape effects~\cite{pouliquen1999scaling}:
\begin{equation}\label{mub-ar}
\mu_b =\mu\left( h,Fr,\textrm{AR} \right) =\tan \theta _1+\frac{\tan \theta _2-\tan \theta _1}{\beta_\textrm{AR}h/\left( Ad\left( Fr+\gamma \right) \right) +1}
\end{equation}
where \(\gamma\) is the intercept in fitting the $Fr$-$h/h_\textrm{stop}$ curves~\cite{forterre2003long} [which is close to zero for all AR in this study according to Fig.~\ref{fig3}(b)] and $\beta_\text{AR}$ is given by Eq.~(\ref{eq4}). Eq.~(\ref{mub-ar}) extends the Pouliquen flow rule by incorporating the influence of particle shape (e.g., AR) on the apparent basal friction of the granular material. This provides a possible way to close the governing equations of granular avalanches consisting of elongated particles and can be used to simulate large-scale avalanche problems. Of course, the relevance of this approach for more complicated particle shapes, such as those encountered in geophysical applications, should be addressed in future research.

\section{\label{Conclusions}Conclusions}
In this paper, we systematically investigate the influence of particle elongation on dense granular flows down a rough inclined plane. By varying the particle aspect ratio AR, we analyze the flow mobility of the granular flow within the framework of Pouliquen's flow rule. We find that elongated particles are more difficult to flow, indicated by a significant rightward shift of the $h_\text{stop}$ curve, but this effect saturates as AR exceeds around $2$. For each individual AR value, Pouliquen's approach collapses the data of $Fr$-$h/h_\text{stop}$ for varying flow thicknesses $h$ and slope angles $\theta$, but a universal collapse for all AR is not achieved; the slope of the $Fr$-$h/h_\text{stop}$ curves, $\beta$, shows an intriguing three-regime dependence on AR. Specifically, $\beta$ remains essentially unchanged for $\text{AR}\lesssim1.3$, followed by a sharp increase for $1.3\lesssim\text{AR}\lesssim2$, and reaches another plateau for $\text{AR}\gtrsim2$. The $\beta$-AR relation is captured by an empirical sigmoidal function and can be used to derive a shape-dependent basal friction law for shallow granular flow modeling.

To understand the three-regime dependence of $\beta$ on AR, we explore microscopic information of the flow, including the statistics of particle orientation and rotational velocity. Particles with elongated shapes tend to form specific arrangements. As AR increases, the degree of particle alignment $S$ increases and the preferred orientation angle $\alpha _{xz}^{p}$ decreases; the latter exhibits a clear transition towards saturation at $\textrm{AR}\approx1.5$, which correlates with the second plateau in $\beta$. On the other hand, the first plateau of the $\beta$-AR relation for small AR values finds its microscopic origin in the rotation statistics of elongated particles. As AR increases slightly over $1$, the rotational ability of the particles, characterized by the slope $k_\omega$ of the $\omega$-$\dot\gamma$ relations, remains nearly unaffected (even though the flow mobility is largely reduced in this regime). Therefore, at least for weakly non-spherical particles ($\text{AR}\lesssim1.3$), particle shape does not introduce fundamental changes of the rotational behaviors of the particles.

Our microscopic analysis above provides possible explanation of how the particle shape controls the three-regime behavior of $\beta$, exploiting different microscopic parameters for different regimes. However, it is stressed that neither particle alignment ($S$ and $\alpha _{xz}^{p}$) nor rotation ($k_\omega$) parameters exhibit a transition at $\textrm{AR}\approx1.3$ that is as sharp as in $\beta$. This seems to suggest a missing micromechanically-based physical quantity which governs the phase transition between flow behaviors of spherical and sufficiently non-spherical particles. Future work should continue seeking this microscopic physical quantity and focus on more particle shape parameters, such as flatness and angularity, to develop a more unified understanding of the role of particle shape in granular flows. 

\begin{acknowledgments}
This research is financially supported by the Open Research Fund Programs of State Key Laboratory of Hydroscience and Engineering (SKLHSE-2023-B-07) and State Key Laboratory of Hydraulics and Mountain River Engineering (SKHL2218). G.G.D.Z. is grateful for the financial support of Sichuan Science and Technology Program (2024NSFJQ0043).
\end{acknowledgments}

\appendix

\section{\label{app:corrugation}EFFECTS OF PARTICLE CORRUGATION FOR CLUMPED SPHERES}
\begin{figure}[htbp]
\includegraphics[width=0.9\textwidth]{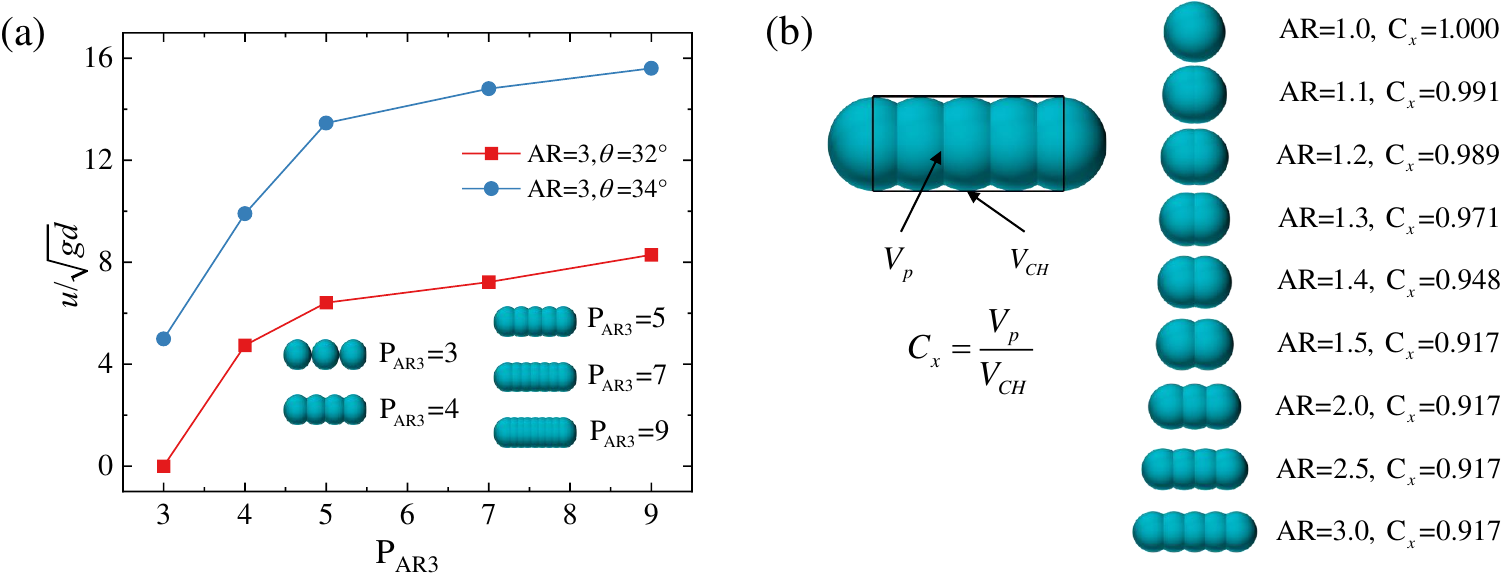}
\caption{\label{fig8}Effects of particle corrugation. (a) Dimensionless velocity $u/\sqrt{gd}$ as a function of $\textrm{P}_\textrm{AR3}$ and (b) convexity $C_x$ of simulated particles. } 
\end{figure}

Using clumped spheres to represent rod-like particles can lead to particle corrugation effects, which serve as an additional frictional mechanism in granular flows \cite{guo2012numerical}. To evaluate the effects of particle corrugation, we define $\textrm{P}_\textrm{AR3}$ as the number of spheres used in each clump for $\textrm{AR}=3$ and run simulations with varying $\textrm{P}_\textrm{AR3}$. As shown in Fig.~\ref{fig8}(a), flow mobility increases with $\textrm{P}_\textrm{AR3}$ due to smoother particle surfaces, but this increase slows down beyond $\textrm{P}_\textrm{AR3}=5$ (our choice). Furthermore, we use convexity $C_x$ \cite{su2020prediction} to quantify the corrugation of our simulated particles, 
\begin{equation}\label{q_convexity}
C_x = \frac{V_p}{V_{CH}}
\end{equation}
where $V_p$ is the volume of the middle part of a clump (excluding two hemispheres at the two ends) and $V_{CH}$ is the volume of the cylindrical convex hull enclosing the same part of the clump, shown by the box in Fig.~\ref{fig8}(b). Note that $C_x$ for spheres is defined to be one. Generally, $C_x$ of our elongated particles is greater than $0.9$ and remains unaltered for $\text{AR} \geqslant 1.5$, which helps minimize the influence of corrugation on our results.

\section{\label{app:domainsize}EFFECTS OF COMPUTATIONAL DOMAIN SIZE}

\begin{figure}[htbp]
\includegraphics[width=0.9\textwidth]{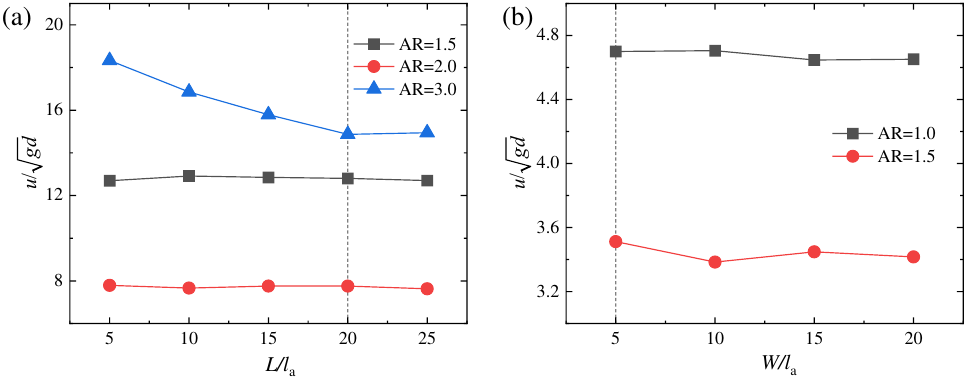}
\caption{\label{fig9}
Effects of computational domain size. (a) Dimensionless velocity \(u/\sqrt{gd}\) vs.\ the domain size in the \(x\)-direction \(L/l_a\) for aspect ratios \(\mathrm{AR}=1.5, 2, 3\) with \(H=20d\) and, respectively, \(\theta=30^\circ, 32^\circ, 35^\circ\). (b) Dimensionless velocity \(u/\sqrt{gd}\) vs.\ the domain size in the \(y\)-direction \(W/l_a\) for aspect ratios $\mathrm{AR}=1$ ($H=20d$, $\theta=25^\circ$) and $\mathrm{AR}=1.5$ ($H=15d$, $\theta=30^\circ$). The dotted lines indicate thresholds beyond which the domain size does not affect the results.}
\end{figure}

To determine the influence of the computational domain size on the results, we vary the dimensions of the domain in the $x$ and $y$ directions. We find that extending the domain length to $20l_a$ in the $x$ direction and $5l_a$ in the $y$ direction can effectively avoid domain size effects (Fig.~\ref{fig9}).

\section{\label{app:parameters}EFFECTS OF MODEL PARAMETERS}
\begin{figure}[htbp]
\includegraphics[width=0.9\textwidth]{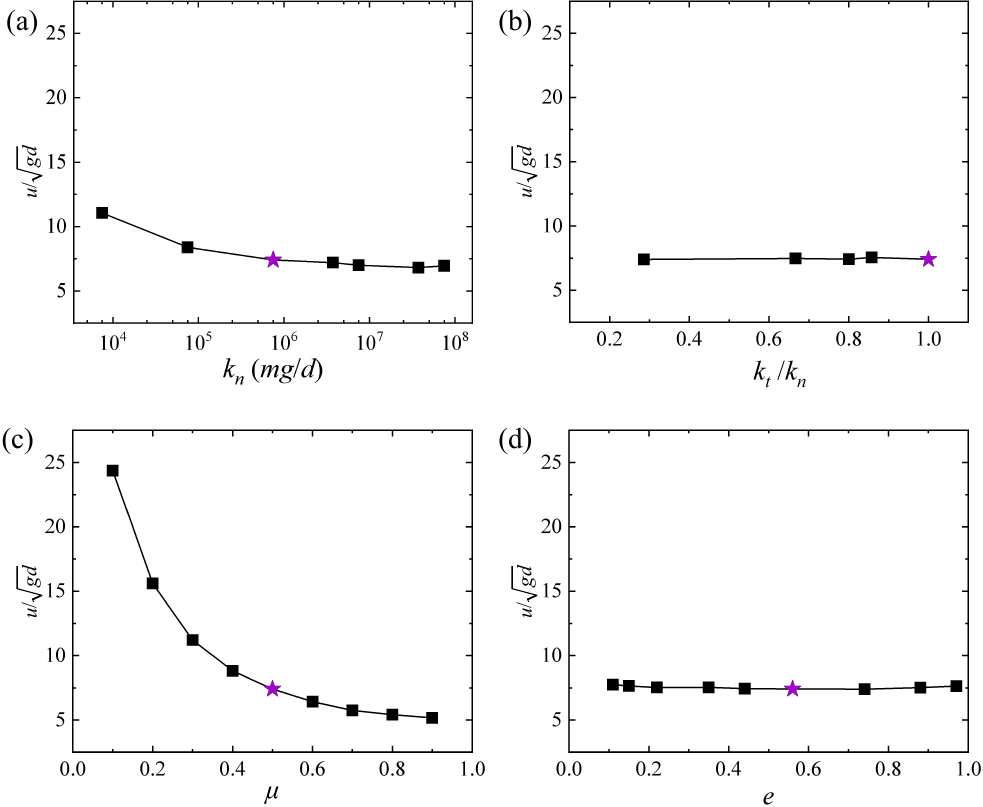}
\caption{\label{fig10}Sensitivity tests for key model parameters. Dimensionless velocity $u/\sqrt{gd}$ as a function of (a) normal stiffness $k_n(mg/d)$, (b) $k_t/k_n$, (c) coefficient of friction $\mu$, and (d) coefficient of restitution $e$. The pentagram symbol indicates the parameters chosen for our simulations. Tested using spherical particles with an initial thickness $H=40d$ and a slope angle $\theta=23^\circ$. } 
\end{figure}

Here we investigate the sensitivity of our results to several key model parameters, including normal contact stiffness, ratio of tangential to normal stiffness, the coefficient of friction, and the coefficient of restitution (Fig.~\ref{fig10}). For an example case of $\theta=23^\circ$ and $H=40d$, we find that the effect of the normal stiffness $k_n$ diminishes as it increases, becoming negligible when it is sufficiently large ($\gtrsim10^6mg/d$) [Fig.~\ref{fig10}(a)]. As for the tangential contact stiffness, in Mindlin’s elastic theory~\cite{mindlin1949compliance} the ratio of the tangential to normal stiffness ${k_t}/{k_n}$ is given by ${k_t}/{k_n} = {2(1 - \nu)}/{(2 - \nu)}$, where $\nu$ is Poisson's ratio. For most elastic materials, Poisson's ratio is in the range of $0\sim0.5$, such that ${k_t}/{k_n}$ falls within the range of $2/3\sim1$. Here we test five different values of ${k_t}/{k_n}$: 2/7 (choice of Silbert et al.~\cite{silbert2001granular}), 2/3, 4/5, 6/7, and 1, corresponding to Poisson's ratios of 5/6, 1/2, 1/3, 1/4, and 0, respectively. It is clear that $k_t$ generally does not affect the results and we set $k_t=k_n$ [Fig.~\ref{fig10}(b)]. 

Figure~\ref{fig10}(c) shows that the results are relatively sensitive to the coefficient of friction $\mu$. Generally, the flow velocity decreases as $\mu$ increases, but the tendency slows down for $\mu\gtrsim0.6$. As we do not aim to reproduce any particular material with $\mu$, we select a typical value of $\mu=0.5$ in the literature. Finally, Fig.~\ref{fig10}(d) shows that the coefficient of restitution $e$ does not affect the results and we choose $e=0.56$ in our simulations.

\section{\label{app:PJFR}IMPROVED POULIQUEN-JENKINS FLOW RULE}
\begin{figure}[htbp]
\includegraphics[width=0.9\textwidth]{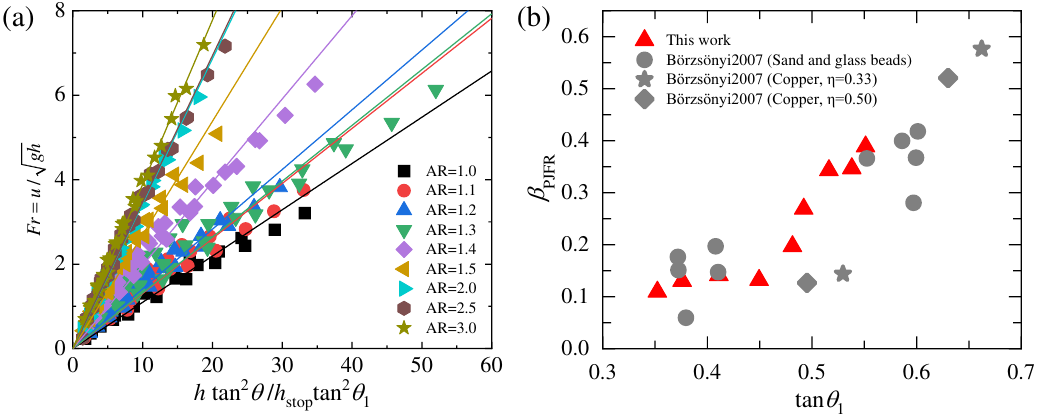}
\caption{\label{fig11}Improved Pouliquen-Jenkins flow rule. (a) Froude number $Fr$ as a function of $h\tan ^2\theta /h_\textrm{stop}\tan ^2\theta _1$. (b) The PJFR slope $\beta_\textrm{PJFR}$ as a function of $\tan \theta_1$. } 
\end{figure}

An alternative framework to analyze our data is the improved Pouliquen-Jenkins Flow Rule (PJFR)~\cite{borzsonyi2007flow}, which considers correlation between the Froude number $u/\sqrt{gh}$ and $h\tan ^2\theta /h_\textrm{stop}\tan ^2\theta _1$. In Fig.~\ref{fig11}(a), our data is recast into this framework and fitted to the following expression:

\begin{equation}\label{eq_pjfr}
Fr=\frac{u}{\sqrt{gh}}=\beta _{\text{PJFR}}\frac{h\tan ^2\theta}{h_{\text{stop}}\tan ^2\theta _1}
\end{equation}
where $\beta_\textrm{PJFR}$ represents the slope of PJFR. Our simulation results agree well with experimental data for glass beads, sand, and copper particles from \citet{borzsonyi2007flow}, as shown in Fig.~\ref{fig11}(b). It can be seen that the slope $\beta_\textrm{PJFR}$ increases with AR [Fig.~\ref{fig11}(a)], similar to the Pouliquen framework, and the $\beta_\textrm{PJFR}$-$\tan \theta_1$ relation shows a very similar trend as previous experimental data [Fig.~\ref{fig11}(b)].

\nocite{*}

%

\end{document}